\documentclass[aps,twocolumn,prl,showpacs]{revtex4}
\usepackage{graphicx,amssymb,amsmath}
\usepackage{graphicx}
\usepackage{dcolumn}
\usepackage{bm}

\newcommand{\Bpar}{$B_{||}$}
\newcommand{\Bperp}{$B_{\perp}$}
\newcommand{\rxx}{$R_{xx}$}
\newcommand{\ryy}{$R_{yy}$}
\newcommand{\rxy}{$R_{xy}$}
\newcommand{\ryx}{$R_{yx}$}

\begin{document}

\title{Tilt-Induced Anisotropic to Isotropic Phase Transition at $\nu = 5/2$}

\author{Jing Xia$^1$, Vaclav Cvicek$^1$, J.P. Eisenstein$^1$, L.N. Pfeiffer$^2$, and K.W. West$^2$}

\affiliation{$^1$Condensed Matter Physics, California Institute of Technology, Pasadena, CA 91125
\\
$^2$Department of Electrical Engineering, Princeton University, Princeton, NJ 08544}

\date{\today}

\begin{abstract} A modest in-plane magnetic field \Bpar\ is sufficient to destroy the fractional quantized Hall states at $\nu = 5/2$ and 7/2 and replace them with anisotropic compressible phases.  Remarkably, we find that at larger \Bpar\ these anisotropic phases can themselves be replaced by isotropic compressible phases reminiscent of the composite fermion fluid at $\nu = 1/2$.  We present strong evidence that this transition is a consequence of the mixing of Landau levels from different electric subbands.   We also report surprising dependences of the energy gaps at $\nu = 5/2$ and 7/3 on the width of the confinement potential. 
\end{abstract}

\maketitle
In the presence of a strong perpendicular magnetic field \Bperp, high purity two-dimensional electron systems exhibit a variety of collective phases \cite{perspectives}.  When the magnetic field is strong enough to force the Fermi level into the $N=0$ lowest orbital Landau level (LL), gapped electron fluids exhibiting the fractional quantized Hall effect (FQHE) appear whenever the ratio $\nu=nh/eB_{\perp}$ of electron density $n$ to the spin-resolved Landau level degeneracy $eB_{\perp}/h$ is one of a large set of odd denominator rational fractions, $e.g.$ $\nu = 1/3$.  For even-denominator fillings, notably $\nu =1/2$ and 3/2, the 2DES remains compressible even as electron-electron interactions create a novel Fermi liquid of composite fermions out of an otherwise dispersionless Landau band. 

At lower magnetic fields, where the Fermi level resides in the $N\geq 2$ excited LLs (and thus $\nu > 4$), the FQHE is not observed. Instead, a family of charge density wave-like states appears \cite{Lilly,Du}.  At half filling of the valence LL, $e.g.$ near $\nu = 9/2$, a strong anisotropy develops in the 2DES longitudinal resistivity at low temperatures, with the ``hard'' and ``easy'' transport directions pinned to orthogonal crystallographic axes.  These anisotropic, or stripe-like compressible phases are believed to be examples of nematic electronic liquid crystals \cite{KFS,MC,FK,AR}.  In the flanks of the same high Landau levels, re-entrant integer quantum Hall effect (RIQHE) states are observed.  These states are believed to be ``bubble'' phases \cite{KFS,MC}, $i.e.$ collective insulators analogous to Wigner crystals but with more than one electron per unit cell.

Between the FQHE-dominated $N=0$ LL and the stripe/bubble phase regime of the $N\geq 2$ LLs lies the $N=1$, or first excited Landau level, where $4>\nu>2$.   FQHE states at $\nu = 7/3$, 8/3, 11/5, etc. recall the famous Laughlin states of the $N=0$ LL.  Several RIQHE states, possibly kin to the bubble phases of the $N\geq 2$ LL, are also observed \cite{JPE2002}.  Most interestingly, FQHE states exist at half filling ($\nu =5/2$ and 7/2), in violation of the odd-denominator rule governing the FQHE in the $N=0$ LL \cite{Willett}. Application of an in-plane component of magnetic field, \Bpar, enriches things still further: The FQHE states at $\nu = 5/2$ and 7/2 are destroyed and are replaced by anisotropic compressible phases strongly resembling the stripe phases of the $N\geq2$ LLs \cite{JPE1988,Lilly2,Pan}.  

We report here surprising new transport phenomena in the $N=1$ LL in tilted magnetic fields \cite{Note1LL}.  First, our experiments reveal that the \Bpar-induced destruction of the $\nu = 5/2$ and 7/2 FQHE states and their replacement by anisotropic stripe-like compressible phases can, under appropriate conditions, be followed, at still larger \Bpar, by the formation of an {\it isotropic} compressible phase resembling the composite fermion metallic states at $\nu = 1/2$ and 3/2 in the lowest LL.  We find that the appearance of this new isotropic phase depends sensitively on the potential confining the 2DES and the relative alignment of Landau levels emanating from the various subbands of that potential.  Second, by comparing samples which differ only in the thickness of the 2DES layer, we find that the energy gaps of the $\nu = 5/2$ and 7/3 FQHE states are larger in the sample with the thicker 2DES.  Remarkably, while the gap for the 5/2 state appears to always fall with \Bpar, the gap at 7/3 can either increase or decrease with \Bpar, depending on the thickness of the 2DES. 

Interest in the 5/2 FQHE state has surged recently owing to the anticipated non-abelian exchange statistics of its quasiparticle excitations and the potential for their application in schemes for topologically-protected quantum computation \cite{DasSarmaRMP}.  Our new findings highlight the complex character of the $\nu = 5/2$ state and suggest that a deep understanding of it has yet to emerge.

Three distinct GaAs/Al$_x$Ga$_{1-x}$As heterostructure samples, A, B, and C, are discussed here.  Each contains a 2DES confined to the lowest subband of a GaAs quantum well symmetrically doped by remote Si impurities deposited in Al$_{0.24}$Ga$_{0.76}$As layers flanking the well.  The width of the quantum well in sample A is 40 nm while in samples B and C it is 30 nm.  The 2DES density and mobility in both samples A and B are $n = 1.6 \times 10^{11}$ cm$^{-2}$ and $\mu = 16 \times 10^6$ cm$^2$/Vs, respectively.  For sample C, $n = 3.0 \times 10^{11}$ cm$^{-2}$ and $\mu = 29 \times 10^6$ cm$^2$/Vs.  Each sample is cleaved into a 5 by 5 mm square and eight InSn ohmic contacts are positioned at the corners and side midpoints of the sample.  The samples are mounted on a rotating stage which allows for {\it in situ} application of both perpendicular and in-plane magnetic fields.  Longitudinal and Hall resistance measurements are performed by driving low frequency ac current (typically 2-5 nA at 13 Hz) between opposing midpoint contacts while recording the voltage difference between corner contacts on the same or opposite sides of the current axis.  \rxx\ is the longitudinal resistance measured with the mean current direction parallel to the in-plane field \Bpar, while for \ryy\ the current is perpendicular to \Bpar. 

\setlength{\abovecaptionskip}{-5pt}
\setlength{\belowcaptionskip}{-5pt}
\begin{figure}
\begin{center}
\includegraphics[width=1.0 \columnwidth] {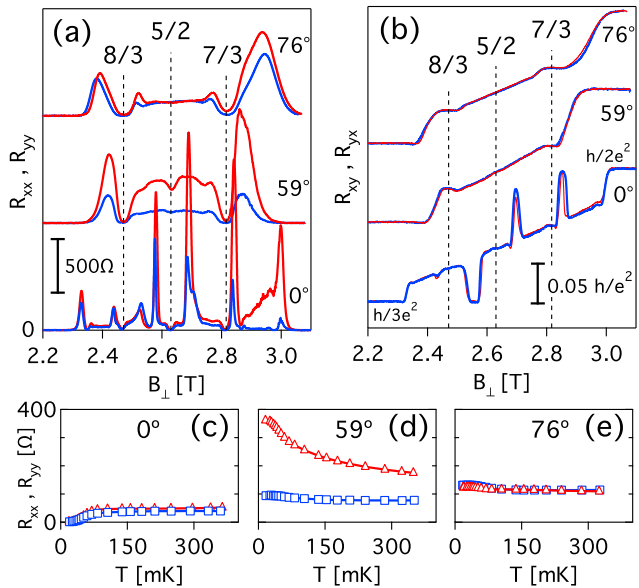}
\end{center}
\caption{(color online) Magnetotransport in the $N = 1$ Landau level in sample A. (a): $R_{yy}$ (blue), $R_{xx}$ (red) and (b): $R_{yx}$ (blue),  $R_{xy}$ (red) at $T \approx$ 15 mK at $\theta = 0^\circ$, 59$^\circ$, and 76$^\circ$.  Traces offset vertically for clarity. (c)-(e): Temperature dependences of  $R_{yy}$ (squares) and $R_{xx}$ (triangles) at $\nu$ = 5/2 at the same tilt angles.}
\label{fig1}
\end{figure} 
Figures \ref{fig1}(a) and \ref{fig1}(b) show longitudinal and Hall resistance data from sample A at various representative tilt angles $\theta$ and $T \approx 15$ mK.  The data are plotted versus \Bperp, and extend over the filling factor range $2<\nu<3$ in the $N =1$ LL. At $\theta = 0$ deep minima in the longitudinal resistance and well-quantized plateaus are observed at $\nu = 5/2$, 8/3, and 7/3.  The expected RIQHE states \cite{JPE2002}, in various stages of development, are also clearly evident in the Hall resistance.  With the exception of the tall peaks in \rxx\ and \ryy\ associated with the RIQHE, which are quite sensitive to sample inhomogeneities, the longitudinal resistance near half filling of the Landau level is isotropic \cite{anomaly}.  Fig. \ref{fig1}(c) shows that at $\theta = 0$ \rxx\ and \ryy\ at $\nu = 5/2$ vanish as $T \rightarrow 0$, as expected owing to the presence of a FQHE energy gap. 

Upon tilting the sample, both the $\nu = 5/2$ FQHE and the RIQHE states are weakened and eventually destroyed \cite{JPE2002,csathy}.  Furthermore, the longitudinal resistance in the vicinity of $\nu = 5/2$ becomes anisotropic, with the ``hard'', or high resistance direction parallel to \Bpar\ as found previously \cite{Pan,Lilly2}.  At $\theta =59^\circ$, the anisotropy in the longitudinal resistance around $\nu = 5/2$ is quite substantial.  In common with the stripe phases at $\nu = 9/2$, 11/2 and other half-filled high Landau levels \cite{Lilly,Du}, the resistive anisotropy at $\nu = 5/2$ and $\theta = 59^\circ$, shown in Fig. \ref{fig1}(d), grows rapidly as $T \rightarrow 0$.  At the same time, deep minima in \rxx\ and \ryy\ and well-quantized plateaus in \rxy\ and \ryx\ signal the continued stability of the $\nu = 7/3$ and 8/3 FQHE states. 

Further tilting of sample A has a profound and unexpected result: the resistive anisotropy near $\nu = 5/2$ disappears entirely and a compressible, non-FQHE phase, dubbed the reentrant isotropic compressible (RIC) phase, appears.  Figures \ref{fig1}(a) and \ref{fig1}(e) show that by $\theta = 76^\circ$ \rxx\ and \ryy\ are essentially identical and temperature independent in the vicinity of $\nu = 5/2$.  Indeed, \rxx\ and \ryy\ remain roughly equal throughout the filling factor range $2<\nu<3$.  At $\nu = 7/3$ and 8/3 robust FQHE states are observed while weak local minima in the resistances near $\nu = 12/5$ and 13/5 suggest nascent FQHE states at these fillings.  Although not shown in Fig. \ref{fig1}, an essentially identical sequence of tilt-induced transitions, from FQHE to anisotropic phase to RIC phase is observed around $\nu = 7/2$ in the upper spin band of the $N=1$ LL. 

\begin{figure} [t]
\includegraphics[width=1.0 \columnwidth]{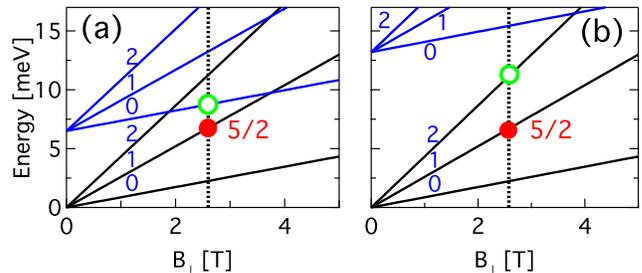}
\caption{(color online) $N=0$, 1 and 2 Landau levels in a perpendicular field emanating from the lowest (black) and first excited (blue) electric subbands in (a) sample A and (b) sample B. Solid and open dots indicate the highest occupied and lowest unoccupied Landau levels at $\nu = 5/2$.  The relative level alignments in sample C are the same as in sample A. Spin splitting of levels ignored.} 
\label{fig2}
\end{figure}

The transport data from sample A at $\theta = 76^\circ$ is reminiscent of that observed at $\theta = 0$ in the $N=0$ LL where the $\nu = 4/3$ and 5/3 FQHE states straddle a compressible phase at $\nu = 3/2$.  This suggests that the results shown in Fig. \ref{fig1} may derive from Landau level mixing.  While electron-electron interactions create LL mixing even in a purely perpendicular magnetic field, the in-plane field component of a tilted field mixes Landau levels and the electric subbands of the confinement potential already at the single particle level.  It is therefore plausible that the relatively small splitting ($\Delta E_{1,0} \approx 6.5$ meV \cite{LDA}) between the ground and first excited subband of the 40 nm-wide quantum well in sample A plays an important role in the phenomena reported here.  Although the first excited subband is $not$ occupied at $B=0$, it is nonetheless true that the lowest unoccupied Landau level (LULL) above the Fermi level at $\nu = 5/2$ is the $N=0$ LL of the first excited subband, not the $N=2$ LL of the ground subband as is usually assumed.  Figure \ref{fig2}(a) illustrates the relative alignment of the first few Landau levels emanating from the two lowest electric subbands of the confinement potential in sample A, with the solid and open dots indicating the highest occupied and lowest unoccupied Landau levels, respectively, at $\nu = 5/2$.  Figure \ref{fig2}(b) shows the more typical alignment in which the LULL is the $N=2$ LL of the ground subband.  Clearly, the distinction between these two alignments depends on both the subband splitting $\Delta E_{1,0}$ and the magnetic field at $\nu = 5/2$.

\begin{figure} [t]
\begin{center}
\includegraphics[width=1.0 \columnwidth]{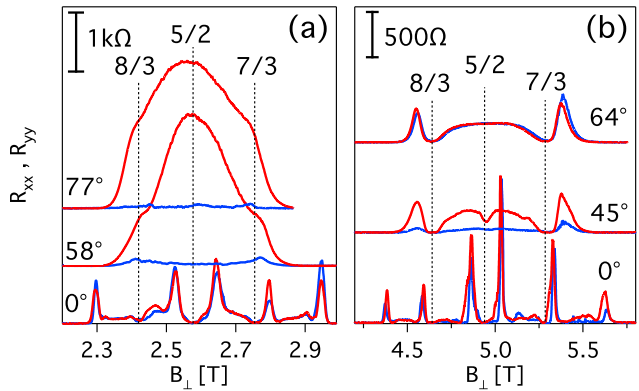}
\end{center}
\caption{(color online) Magnetoresistance at three representative tilt angles in (a) sample B  and (b) sample C at $T \approx$ 15 mK near $\nu$ = 5/2.}
\label{fig3}
\end{figure}

To test the importance of Landau level and subband mixing we turn to data from samples B and C, both of which confine the 2DES in a 30 nm quantum well.  Figure \ref{fig3} reveals that transport around $\nu = 5/2$ evolves very differently with tilt in these two samples. In sample B, which has a narrower quantum well but the same 2DES density and mobility as sample A, the destruction of the 5/2 FQHE with tilt is followed by the development of a very large transport anisotropy.  This anisotropy persists out to the largest tilt angles studied ($\theta = 77^\circ$); no RIC phase is observed.  In contrast, the behavior of sample C is quite similar to sample A.  The 5/2 FQHE is destroyed by small tilt and an anisotropic state appears.  At large tilt angle the anisotropic state disappears and a RIC phase emerges.  As in sample A, the tilt dependence of the $\nu = 7/2$ state in samples B and C is qualitatively identical to that of the 5/2 state. 

The calculated subband splittings in samples B and C are $\Delta E_{1,0} = 13.2$ and 11.4 meV respectively \cite{LDA}; the slight difference being due to the different electron densities in these 30 nm quantum well samples.  These subband splittings are roughly twice that of sample A.  Consequently, for sample B, which has the same electron density ($n = 1.6 \times 10^{11}$ cm$^{-2}$) as sample A, the LULL at $\nu = 5/2$ is the $N=2$ LL of the ground subband, as indicated in Fig. \ref{fig2}(b). In contrast, in sample C, with its higher electron density ($n = 3.0 \times 10^{11}$ cm$^{-2}$), the alignment reverts qualitatively to that of sample A as shown in Fig. \ref{fig2}(a).  These observations suggest that the appearance, at high tilt angle, of a RIC phase at $\nu = 5/2$ depends critically on the LULL being the $N=0$ LL of the first excited subband.  If instead, the LULL is the $N=2$ LL of the ground subband, as in sample B and depicted in Fig. \ref{fig2}(b), the 5/2 state becomes increasingly anisotropic out to the highest tilt angles studied.

A qualitative, if highly oversimplified understanding of the distinction between the two types of tilted field behavior at $\nu = 5/2$ is readily constructed.  If, as in samples A and C, the LULL is the $N=0$ LL of the first excited electric subband, then tilt-induced mixing of this level into the $N=1$ LL of the ground subband from which the 5/2 state originates, should give electron correlations some of the character of the $N=0$ LL.  Since the $\nu = 1/2$ and 3/2 states in the $N=0$ LL are compressible composite fermion metals, it is then not surprising that the highly tilted 5/2 state in samples A and C resembles them.  In contrast, if, as in sample B, the LULL is the $N=2$ LL of the ground subband and mixing to this level is enhanced by tilting, it is perhaps not surprising that the 5/2 state increasingly resembles the anisotropic stripe phases found at $\nu = 9/2$, 11/2, $etc.$ in the $N=2$ LL.  While attractive, we emphasize the very qualitative nature of this scenario.  At the high tilt angles studied here, mixing with several LLs may be important \cite{trilayer}. Furthermore, we note that the present results were obtained using symmetrically-doped quantum well samples.  Whether the symmetry of the confining potential is relevant remains to be investigated.  Finally, the spin degree of freedom has been ignored.  However, the qualitatively identical tilted field evolution of the 5/2 and 7/2 states that we observe in all three samples suggests that spin, while obviously resolved, plays little role in the phenomena reported here.

\begin{figure}
\begin{center}
\includegraphics[width=1.0 \columnwidth]{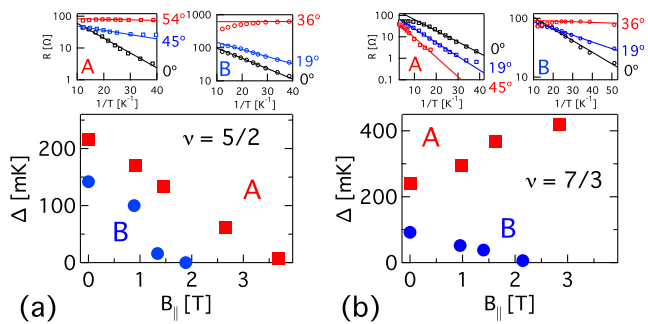}
\end{center}
\caption{(color online) Activation gaps at (a) $\nu = 5/2$ and (b) $\nu = 7/3$ vs. \Bpar\ for samples A (squares) and B (dots) along with selected Arrhenius plots of the longitudinal resistances.} 
\label{fig4}
\end{figure}
We now turn to our results on the energy gaps $\Delta$ of the fractional quantized Hall states at $\nu = 5/2$ and 7/3 for small tilt angles.  Fig. \ref{fig4} displays $\Delta_{5/2}$ and $\Delta_{7/3}$ $vs.$ \Bpar\ for samples A and B, as determined by fitting $R_{yy}$ to the standard form $exp(-\Delta/2 k_{B}T) $.  As has been previously observed \cite{JPE1988}, the gap $\Delta_{5/2}$ declines rapidly with tilt in both samples.  While this suppression of $\Delta_{5/2}$ may result from the increased spin Zeeman energy that tilting produces, it might instead be due to competition between the 5/2 FQHE phase and a compressible stripe phase \cite{haldane_rezayi}.  What is remarkable here is that the 5/2 gap in sample A is significantly larger than in sample B.  These two samples have essentially identical electron densities and mobilities, but have different quantum well widths (40 nm in sample A, 30 nm in sample B).  That the gap is larger in the wider well sample contrasts with the usual expectation that the thickness-induced softening the Coulomb interaction weakens FQHE phenomena.  Recent theoretical work \cite{peterson} suggests however that wider well widths can in fact help stabilize the 5/2 FQHE state, at least when it is approximated by the Moore-Read \cite{moore_read} Pfaffian wavefunction.

The tilted field behavior of the gap at $\nu = 7/3$ in samples A and B is even more surprising.  In addition to $\Delta_{7/3}$ at zero tilt being larger in the wider well sample, the dependence on \Bpar\ is opposite in the two cases.  As previously observed \cite{dean}, $\Delta_{7/3}$ initially rises with \Bpar\ in the 40 nm sample \cite{larger_tilt}. While the origin of this increase is unknown, it may imply that spin-reversed excitations dominate in sample A.  However, as Fig. \ref{fig4} demonstrates, $\Delta_{7/3}$ $falls$ with increasing tilt in sample B.  That such different behavior develops immediately upon tilting points to a qualitative distinction, already at \Bpar=0, between the quasiparticle excitations and/or the ground state at $\nu = 7/3$ in these two samples.  (A very similar distinction in the tilted field behavior of the $\nu = 8/3$ FQHE in samples A and B is also observed.)  Although an understanding of these results is so far lacking, it must eventually be rooted in the only significant difference in the two samples: the width of their quantum wells.

In conclusion, we have reported several new correlation phenomena in 2D electron systems in which the Fermi level lies in the $N=1$ Landau level. At $\nu = 5/2$ and 7/2 we find that the anisotropic compressible phases which initially replace the incompressible fractional quantized Hall states upon application of an in-plane magnetic field can themselves be supplanted at larger in-plane fields by a new isotropic compressible (RIC) phase.  We have presented strong evidence that Landau level and electric subband mixing is crucial to the transition to this new RIC phase. In addition, we have shown that the details of the potential confining a 2DES can qualitatively influence both the strength and tilted field behavior of the fractional quantized Hall states at $\nu = 5/2$ and 7/3. 

We thank Edward Rezayi for helpful discussions.  This work was supported via Microsoft Project Q.


\begin{references}
\bibitem{Note1LL} As we shall discuss, tilted magnetic fields mix Landau levels and electric subbands.  We use the identifier ``N = 1 LL'' only to indicate the Landau level from which specific filling factors, $e.g.$ $\nu = 5/2$, originate at zero tilt.
\bibitem{anomaly} At $\theta = 0$ sample A exhibits a pronounced difference between \rxx\ and \ryy\ for $2 \lesssim \nu \lesssim 2.3$.  This feature is not consistently found in comparable samples and we attribute it to irrelevant extrinsic effects.
\bibitem{LDA} $\Delta E_{1,0}$ is estimated via self-consistent solution of the Schroedinger and Poisson equations in the heterostructure potential in the local density approximation. The standard Hedin-Lundqvist potential \cite{Hedin} is used to approximate exchange and correlation effects.  Our numerical methods successfully reproduce the results of Stern and Das Sarma \cite{SdS} where appropriate.
\bibitem{larger_tilt} We reserve for later publication the behavior of the 7/3 state at larger tilt angles.
\end{references}
\end{document}